\documentclass[useAMS,usenatbib]{mn2e}
\usepackage{color}

\newcommand{\highlight}{\color{red} }
\renewcommand{\highlight}{}
\newcommand{\be}[1]{\begin{equation} \label{eq:#1}}
\newcommand{\ee}{\end{equation}}

\newcommand{\solrad}{\ifmmode{R}_{\rm S}\else${R}_{\rm S}$\fi}
\newcommand{\solmas}{\ifmmode{M}_{\rm S}\else${M}_{\rm S}$\fi}
\newcommand\lta { \mathrel {\hbox to 0pt {\lower 3.7pt \hbox{$\sim$}
      \hss} \raise 1.7pt \hbox{$<$}}}
\newcommand\gta { \mathrel {\hbox to 0pt {\lower 3.7pt \hbox{$\sim$}
      \hss} \raise 1.7pt \hbox{$>$}}}

\newcommand\name{LAMBERT}
\renewcommand\name{GeoSphere}

\newcommand\tableone{
\begin{table}
\label{tab:props}
\begin{center}
 \caption{Conditions for a 4m Geostationary sphere}
 \begin{tabular}{ll}     
 \hline
 \hline
Parameter & value \\
 \hline
\\
\multicolumn{2}{l}{Environmental conditions at \name{}}\\
Geocentric distance $r_o$ km & 42,164\\
Solar $m_V$  & $-26.74$ \\
Earth $m_V$ ($\beta=90^\circ$,quadrature) & $-20.6$ \\
Earth $m_V$ ($\beta=20^\circ$) & $-16.0$ \\
Moon $m_V$ & $\gta -12.74$\\
\\
\multicolumn{2}{l}{Physical and observable \name{} properties}\\
Adopted sphere diameter $d_s$ m & 4 \\
Angular diameter arcsec & 0.0196\\
Adopted albedo $a$ & 0.9 \\
$q$ (phase law parameter) & 1.5 \\
$m_V$ (opposition) & 10.08\\
$m_V$ (quadrature) & 11.32\\
 \hline
\end{tabular}
\end{center}
Notes: ``Lambert's law'' for scattered radiation is assumed, 
corresponding to a perfectly diffusing sphere. Magnitudes
were computed assuming the distance to the sphere is 
$r_o-r_\oplus$.  $\beta$ is the angle between the geocenter-\name
line and geocenter-Sun line. 
\end{table}}

\title[\name]{Century-Long Monitoring of Solar Irradiance and Earth's Albedo 
Using a Stable Scattering Target in Space}
\author[Philip Judge and Ricky Egeland]
{Philip G. Judge\thanks{Email: judge@ucar.edu} and Ricky Egeland\thanks{Also affiliated with Montana State University}\\
High Altitude Observatory,
       National Center for Atmospheric Research\thanks{The National %
       Center for Atmospheric Research is sponsored by the %
       National Science Foundation}\\
       P.O.~Box 3000, Boulder CO~80307-3000, USA}
\begin{document}

\date{Draft v6.1 3 Nov 2014}

\pagerange{\pageref{firstpage}--\pageref{lastpage}} \pubyear{2014}

\maketitle

\label{firstpage}

%
%

\begin{abstract}
An inert sphere of a few meters diameter, placed in a special stable
geosynchronous orbit {\em in perpetuo}, 
 can be used for a variety of scientific
experiments.  
Ground-based observations of such
a sphere, ``\name{}'', can resolve very difficult problems in
measuring the long-term solar irradiance.  \name{} measurements will
also help us understand the evolution of Earth's albedo and 
climate over at least the next
century.  
\end{abstract}

\begin{keywords}
Albedo, Planetary Radiation, Solar Activity, Solar Radiation, Spectrophotometry, Variability
\end{keywords}

\section{Motivation}

The Sun is a very stable object.  {\highlight The Sun's
``irradiance'' (bolometric flux measured since 1978 from above the earth's
atmosphere) varies by 0.06-0.1\%{} peak-to-peak, on time scales of a
11 years \citep[see, for example, recent reviews by][]{Frohlich2013, Willson2014}.  }
This variation follows the well known
``sunspot cycle'' \citep{Schwabe1844}. On shorter time scales, 
the evolution and disk passage of sunspot groups
produce variations that are often 0.2\%.
Given that sunspots are magnetic \citep{Hale1908b}, these variations
arise because of a variable solar magnetic field. While of 
small amplitude, these changes are
far more rapid than expected from first principles. They are believed
to be generated by a magneto-hydrodynamic dynamo, drawing energy from
differential rotation and turbulence in the Sun's interior. The dynamo 
engine changes the Sun from a benign object, stable over time scales of 
100,000 years and radiating essentially as a black body near 6000K, to 
a rapidly varying object.{\highlight The dynamo also causes far more variable
high energy UV and X-ray photons \citep{White1977},} 
wind variations, and intermittent
energetic particles and magnetized plasma clouds.  

The earth's climate is sensitive to several factors,
the Sun is by far the largest external factor.  Herein lies a problem.  We 
have no physical explanation of the solar dynamo, thus our
understanding of irradiance changes remains empirical. Nor do we have 
accurate measurements of solar variability on time scales longer than decades.  
Long-term variability is a contentious and active 
research area, plagued by the absence of hard data  
\citep[e.g.][]{Judge+Others2012}. 
Irradiances have been measured from 1978 using radiometers in
space \citep[e.g.][]{Willson2014}.
Pre-flight calibration precisions of
$\approx$500 parts per million (ppm), 0.05\%, have been achieved.  But 
independent data from different space missions have failed to provide 
the needed long-term stability.   
First, the
highly precise irradiances measured by different experiments differ by 0.5-1\%
\citep[see figure 1 of][]{Willson2014}, i.e. disagreements of 10 to 20$\sigma$,
on timescales longer than 11 years.
These systematic thwart
attempts to measure variations longer than one satellite lifetime. 
Second, instrumental sensitivities change
over mission lifetimes which are, unfortunately, 
rather close to 11 years. Third, expensive new radiometers 
must be regularly launched.

We propose an alternative solution that can provide society with solar
brightness measurements that are stable over centuries at a time when
climate change is of first importance.  
Our long-term precision photometric {\em goal}
is set to 0.01\% averaged over a week. This may not be 
achievable, thus 
our less stringent 
{\em requirement} is a 0.1\% precision over a century, which will
uniquely constrain long-term solar variations at a level still 
of great interest
 \citet{Shapiro+others2011,Lubin+others2012}.
We adopt
high precision differential photometry of a target in space
against stable ensembles of stars.   Nightly precisions of stellar observations of
``only'' 0.1\% are achieved with existing small telescopes
\citep{Young+others1991,Henry1999}, with 
seasonal averages precise to 0.01\%.  

\section{A simple solution}

Photometry of planets,
asteroids and the moon present difficulties when trying to
extract solar variations near 0.01\%.  Planets have atmosphere-related
albedo variations (for Neptune, $m_V \approx 8$, they are $\approx
4$\%, \citealp{Lockwood+others1991}).  Asteroids have variable
brightnesses dependent on their rotational aspect and highly variable
geocentric distances.  The moon is so bright and extended, with diverse 
phase-dependent brightnesses of its features, that accurate relative
photometry with standard stars is not possible.

We suggest that a diffuse object, ``\name{}'', be placed in a
stable geosynchronous orbit (GSO).  
{\highlight The need for a high level of geometric symmetry argues strongly for an
unemcumbered sphere.  }
A sphere of a few
meters diameter has $m_V\approx 10$ at opposition independent of
orientation, and
the optical appearance of a star (Table~1. Placed at L$_2$, $m_V\approx 18$). 
Photometric observations of \name{}
with ground-based telescopes can in 
principle provide century-long monitoring of \name{}'s brightness at
0.01\% precision \citep{Young+others1991,Henry1999} using statistical
averages of many ($\approx 20+$), nightly observations.   
If the scattering properties of the sphere in orbit can 
be measured, calibrated and/or modeled at the same level of precision, 
solar brightness variations will follow.  
\name{} should spin with a period of a few seconds, less than
a typical integration time of precision photometry (a minute).  

Table~1 lists conditions for a 4m sphere as seen from Earth.
We adopt a 4m diameter only because fairings of large launch vehicles 
extend to 5m. Smaller vehicles might be used 
with on-orbit expansion of a pre-compacted sphere (e.g., a ``Hoberman sphere''). 
Using data from \cite{Allen1973}, the
Johnson $m_V$-magnitude when the sphere is at angle $\alpha$ from the
anti-Sun point, is, with radius $r_s=d_s/2$:
\be{v} m_V(\alpha) \approx - 26.74 - 2.5
\log_{10} \left (\frac { r_s^2 }{(r_o-r_\oplus)^2} \frac{a}{q}
\phi(\alpha) \right).  \ee 
Here we use a perfectly diffusing sphere
whose phase function $\phi(\alpha)$  obeys Lambert's law: 
\be{lambert} \phi(\alpha) = \left [ \sin
  \alpha + (\pi-\alpha) \cos \alpha) \right ] / \pi 
\ee 
for which $q=1.5$.  Observing \name{} close to
opposition gives $\phi(\alpha \ll 1)\approx1$, and  $m_V = 10.08$.

\citet[][their table 1]{Young+others1991} calculate 
exposure times
needed to achieve 0.05\% RMS photon noise
in the Str\"omgren $b$ and $y$ filters 
at 4670 and 5470 \AA{}, with 
widths of 160 and 240 \AA{} respectively.
Assuming $m_V(\name)=10.08$, 
an integration time of $\lta 100$s, we find that 
telescope aperture $D$ and \name{} diameter $d_s$ must satisfy:
\begin{eqnarray}
(d_s D)^2 &\gta &36 {\rm \ m^4} \ \ \ {\rm (Stromgren}~by~{\rm at~opposition).}
\end{eqnarray}
In the general case of a spectral sampling of $w$ \AA, 
\be{spectra} \nonumber
(d_s D)^2 \gta 5800/w {\rm \ m^4}. \ \ \ 
\ee
For observations with the $w=10$\AA{} sampling of the SORCE-SIM
instrument \citep{Harder+others2000} for example, then $d_s=4$m and 
$D\gta 6$m.

\tableone

\section{Solar brightness observations}

Narrow-band photometric measurements of solar-like stars are routinely
made using small ($\lta0.8$) Automated Photoelectric Telescopes (APTs,
\citealp{Henry1999, Hall+others2009}).  In practice, combining all
sources of error, those brighter than $m_V=8$ have a daily rms
precision of 1 milli-magnitude in $b$ and $y$.  Scaling to the
brightness of \name{}, telescopes of 1.5 m diameter are needed to
measure the brightness of a 4m diameter \name{} at the same daily
level of precision.  With telescope time devoted exclusively to
\name{}, the precision will improve over the 0.1\% nightly means
reported in the stellar APT programs.

However, qualitatively different observations must be made to permit
differential photometry of \name{} against stars.  Objects in GSO
are observed most simply by switching off a telescope's
astronomical drive.  In time $t$ the Earth's rotation sweeps out
$15\arcsec\cos\delta~t$ of sky, where $\delta$ is the object's
declination.  As a result, background stars passing near to \name{}
will contribute to the integrated flux.  Consider a photometer fed
from an aperture in the telescope's focal plane of $5\arcsec$
diameter, $\delta =0^\circ$, and $t= 100$ seconds.  The area covered
is $6\times10^{-4}$ square degrees. If the average number of stars
with visual magnitude $>m$ is $N(m,b)$ per square degree at Galactic
longitude $b$, then the number of stars crossing the field per 100s
observation is
$$
n= 6\times10^{-4} N(m_V,b). 
$$
Using data from \cite[][\S 117]{Allen1973}, for $m_V=10$, $\log_{10}
N(m_V,b)$ varies between 1.25 (Galactic latitude $b=0^\circ$) and 0.55
($b=90^\circ$); for $m_V=20$ we find $\log_{10}N(m_V,b=0,90)=5.0$ and
3.4 respectively.  For $b=0$ (worst case), 60 stars of magnitude 20 or
brighter will cross the aperture; for $m_V \lta10$, on average 1 in
100 such integrations will have one star of brightness similar to
\name{} affecting the integrations.  One solution might be to feed
three mutually calibrated photometers from the focal plane of the
telescope, offset along a line of constant declination.  With \name{}
measured from one part of the focal plane, the same stars can be
measured within a second before and after the target in the other
photometers, allowing subtraction of their contribution.

\section{Earth albedo variations }

Changes in Earth's albedo might be monitored observing  \name{} 
close to quadrature, where it is illuminated by the Earth's 
sunlit hemisphere. Then $\alpha_0 \lta \pi/2$,
$\phi(\alpha_0) \lta 1/\pi$ and the brightness of \name{} from direct
sunlight is $m_V\gta 10.96$.  Using mean earth albedo data 
\citep{Allen1973}, scattered light from the Earth adds
$\approx 0.4\%$ of the direct solar
flux to the \name{} at quadrature.    By measuring \name{}'s differential 
brightness as it crosses the sky, with statistical averaging, 
the goal of 
of 0.01\% precision might be achieved. Long-term relative changes in
Earth's albedo greater than  2.5\% should be measurable by \name.  

\section{Technical challenges}

\subsection{Orbital stability}

The orbits of GSO satellites are perturbed by 
non-spherical components of the Earth's gravitational field, by lunar
influences and by momentum transfer from solar radiation.  Even so, 
certain orbits are stable enough over decades, for
the purposes discussed here.  Orbital drift through the
Poynting-Robertson effect is unimportant for conceivable area-to-mass
ratios for a sphere.  Quoting from the conclusions of
\cite{Friesen+others1992}:

\begin{quote}
``Orbits of large satellites are radially stable throughout the region
within 2000 km above and below the geosynchronous radius.  Satellites
placed in such orbits can be expected to go no more than 50 km above
their original apogees or more than 50 km below their perigees for at
least several centuries, barring collisions.... `A stable plane'
exists at the GSO distance.. for which orbit plane motion is quite
limited... if the initial angular momentum vector of a circular GSO
is aligned with the 7.3$^\circ$ displaced axis about which
precession takes place for the initially equatorial case, orbit plane
motion will be limited to 1.2$^\circ$ of the initial plane.''
\end{quote}

A 50 km change in orbital radius gives a 1 mmag change in brightness. 
These theoretical considerations are borne out in part by measurements
of the orbits of abandoned satellites.  
INTELSAT 1, abandoned on December 2
1969, has remained in an orbit with
semi-diameter between 42,130 and 42,200 km with eccentricity between 0
and 1.6$\times10^{-3}$ \citep[e.g.][]{Ulivieri+others2013}
for over 41 years. 
There are two longitudes of orbital stability arising from
higher order terms in the dynamic equations \citep{Friesen+others1992},
105W and 75E, running through western North America and the middle
of Asia respectively.   These would be the desired orbits for \name.

\subsection{Collision hazard}

\cite{Friesen+others1992}  compute the ``collision hazard'' for orbits of
small inclination (closing speed of two objects in GSO being
$v_c \sin i$ with $v_c$ the orbital speed of 3 km~s$^{-1}$):
\be{cr}
CH = \frac{v_c\sigma}{2\pi^2R^2\Delta R}  \ \ \ {\rm s^{-1}}, 
\ee
where $\sigma \approx \pi (d_s/2)^2$ is the geometric cross section of
the colliding objects, $R$ is the GSO radius and
$\Delta R$ the rms altitude range of the object, $\sim 50$ km.  Per
impacting object this becomes 1 impact in $10^9$ years.  There are only of
order 400 satellites in GSO  around earth, and $\lta$ 300 
other small detected objects large enough for concern
\citep[][Figure 2.2]{nsa_damage}. 
The largest
threat to \name{} is from smaller natural micrometeoroids.

Conversely, a non-powered \name{} represents a collision hazard to
other spacecraft in GSO.  The International Telecommunication Union 
(ITU) has mandated an
end-of-life strategy for GSO objects, with ability to move
out of the GSO region a requirement for new satellites. The
science goals of \name{} are to remain in orbit for perpetuity,
with no pre-planned end-of-life anticipated.   
The ``stable plane'' solutions found by
\citet{Friesen+others1992} minimize relative speeds between objects in
these orbits, and reduce space debris by reducing both impact rate and
debris scatter. If these orbits are adopted as ``graveyard orbits'',
there will be no ITU problem.

\subsection{Secular changes in scattering properties}

The GSO space environment is not ``clean''.  {\highlight Surfaces exposed to
propellant, silicone, and/or exposed to the solar UV flux and
ionizing radiation, change their reflective properties
\citep{Dever+others2012}.  Oxidation is a serious issue in low earth orbit (LEO) where 
oxygen densities are 
$10^{5-6}$ cm$^{-3}$ in typical models, but oxygen 
concentrations in GSO are some $10^5$ times smaller. 
Instead, in GSO, high energy particles and photons are the larger concern. 
For example, the 
absorptance (1 minus albedo) of white paints 
commonly used on spacecraft increases by factors of 1.3 to 2.8
after $\approx$ 100 days exposure to solar-like UV and
energetic electrons in LEO \citep[][Table 4]{Dever+others2012}.  To achieve 
an albedo stability over a century, \name{}'s
surface should be a simple metal, alloy, or crystalline 
material not susceptible to such changes, like telescope optics 
in space.   Such substances would need 
testing before launch, and perhaps 
``pre-aged'' by deep exposure to energetic ions, electrons and photons. }

All objects in GSO develop craters, punctures and chemical
contamination from micrometeoroid (MM) impact. Man-made orbital debris
is less important in GSO orbits and is neglected here
\citep{nsa_damage}.  We consider particles mostly in excess of 1
micron size, those that are smaller being expected to exhibit unbound
orbits because of the strong solar radiation pressure.

 As damage accumulates the phase function of \name{}
 will evolve from, for example, Lambertian to cratered, or lunar-like.
 Diverse datasets collected from extended space missions indicate the
 effects of MM impacts on surfaces.  The large solar cells on space
 station MIR indicated an accumulated surface coverage of MM damage of
 0.045\% in 10 years of exposure in LEO \citep{Smirnov+others2000}.
 Measurements in GSO are rarer. But the gravitational focusing
 for typical MMs is modest ($v \sim 20 $ km~s$^{-1}$) compared with Earth
 escape speed ($v_{\rm esc} = 11-4$ km~s$^{-1}$ in LEO and GSO
 respectively) and so fluxes will be similar.  \name{} will accumulate
 MM surface damage that covers $\approx0.45\%$ per century, a fractional 
rate 
$k= 4.5\times10^{-5}$ y$^{-1}$.  Like the
 moon, cratering enhances 180$^\circ$ scattering at the expense of
 scattering at smaller angles.  Changes to the geometric albedo may be
 smaller, depending on the surface material used and the surface 
optical contamination. {\highlight Spill-over of impact debris beyond the craters'
areas should be negligible if electrical potentials
of attraction (polarization or net charge) are less than the kinetic
energy of any charged exploding debris ejected outwards from the sphere.}

Damage of 0.45\% in area translates to smaller than 0.45\% 
changes in scattered light.  
The albedo after time $t$ in years is
\be{change}
a_t = (1 - kt) a_0 + (kt) b = a_0 - kt(a_0-b)
\ee
where $a_0$ is the initial albedo and $b$ the albedo of damaged
areas.  Since $b > 0$ MM-induced degradation in albedo occurs slower than
$(1-kt) a_0$.  Nevertheless, it should be mitigated, modeled and/or
measured. One solution might be to expose \name{} to cosmic-like dust
particles to mimic a sufficiently long exposure ($>k^{-1}$ y) in
geosynchronous orbit, such that the phase function remains constant.
A hypervelocity gun\footnote{For example, The White Sands Test
  Facility Remote Hypervelocity Test Laboratory, {\tt
    http://www.nasa.gov/\-centers/wstf/laboratories/hypervelocity}.}
might be used.  Another might be to calibrate the scattering
properties of \name{} using future ground-based stable laser sources.

\subsection{Comparison Stars}

To achieve long term precisions
of fractions of milli-magnitudes, a \name{} observing program would
have to devote effort to identifying suitably non-varying comparison stars.
This is not a trivial task
\citep{Henry1999, Hall+others2009}.
Geosynchronous objects circumnavigate the celestial equator
 once per day, the right ascension at opposition
drifting about 1 degree/day. \name{} would therefore require a
significant number of cross-calibrated comparison stars in order to
place its differential magnitude on a uniform scale throughout the
year.  Typical separations within photometric standard groups of three
stars are near $3^\circ$ and always 
less than 13$^\circ$ \citep{Hall+others2007b,Lockwood+others2007}.
As a rough estimate, 
for a maximum intra-group separation of 4$^\circ$, we find that $345$
highly stable stars distributed along the celestial equator would need
to be identified, with magnitudes near that of \name{}.
                                                                               
\cite{Henry1999} found that stars of spectral class F0-F3 and A8-A9
have rms probabilities of long-term (seasonal) stability better than
$0.0005$ mag of 63.3\% and 87.5\% respectively, and that short-term
photometric stability tends to correlate with long-term stability.
These findings should be used in any comparison star search.  The
observing program of the current Kepler K2 Mission schedules two
$\approx75$ day pointings falling within $\pm 10$ degrees of the
celestial equator, and will obtain  precise photometric lightcurves for
thousands of stars covering about $10\%$ of the region of interest.
The planned \textit{Transiting Exoplanet
  Survey Satellite} (TESS) will obtain precise photometric timeseries
of 27-day duratation for bright FGK-type stars in the entire
equatorial region.  These surveys will allow refinement of a list of
candidates for \name{} comparison stars.

\subsection{Other Problems}

In GSO, thermally-induced geometric changes are dominated by
eclipses at equinoxes.  We find such 
orbital changes to be important at the level of 0.001\% only
for a few minutes as \name{} enters and exits the Earth's shadow.

\name{} will be measured only when out of eclipse ($\alpha > 10.9^\circ$) not
at opposition.  A thin earth crescent is always visible.  
Table~1 shows that when separations $\beta$ of the 
Sun and Earth are $20^\circ$, the 
using Lambertian scattering, the Earth is 10.8 magnitudes dimmer than the Sun
(a flux ratio of $2\times10^4$).  While below our target precision, this extra light 
is something to be considered in analyzing \name{} data.  Most likely one will solve for
the Sun's brightness and suitable parameterizations of albedo and phase functions 
of \name{} and Earth in the analysis.

For a GSO, \name{} observations are best made within a
few hours of midnight, daily.  Irradiance variation data show significant
power (excluding flares which can be identified and removed)
on time scales of hours and days, that is related in both phase and 
amplitude to the decadal variations.  
Using hourly irradiance data from the VIRGO 
instrument \citep{Virgo1997}, we find that sampling every 3 days 
suffices to reduce high frequency aliasing to 10\% of the total variance. 

Lastly, care should be taken to determine the polarization properties of
a \name{} which is always irradiated anisotropically.   This can be determined 
prior to flight and using laboratory models, as well as measured directly 
during from telescopes on the ground.  

\section{Conclusions}

We have suggested that an inert scattering sphere be placed in a
very stable geosynchronous orbit in order to make measurements of the solar flux
variations, and measurements of the Earth's changing albedo.  There
are technical challenges but no fundamental show-stoppers 
in making these measurements. To achieve a
precision better than $0.1$\% 
over centuries, micrometeoroid-induced surface degradation must
be mitigated, modeled and/or measured.  Such precisions are of
great interest to both solar and terrestrial climate change physics.

There is perhaps a direct conflict with ITU's requirement for exit
strategies on GSO spacecraft.  But if the community adopts the
``stable plane'' orbits identified by \citet{Friesen+others1992} 
as ``graveyard orbits'', the ITU mandate will not be applicable
with a \name{} placed in such an orbit.

{
Finally, one year of $ub$ photometric observations of \name{} 
should reveal phase relationships in spectral irradiance changes above and below 400nm, 
 resolving a debate of importance to climate science \citep{Harder+others2009}.}

\section*{Acknowledgments}

We are grateful for discussions with Giuliana de Toma, Roberto Casini, 
and a conversation with Aaron Rosengren, James Garry and Duncan Smith via
ResearchGate.  {\highlight The referee is thanked for very useful suggestions. }

\label{lastpage}

\end{document}